\documentclass[letterpaper]{JHEP3}

\usepackage{amsmath,amsthm, amssymb}
\usepackage{epsfig,multicol}

\newcommand{\be}{\begin{equation}}
\newcommand{\ee}{\end{equation}}
\newcommand{\ba}{\begin{array}}
\newcommand{\ea}{\end{array}}
\newcommand{\bea}{\begin{eqnarray}}
\newcommand{\eea}{\end{eqnarray}}
\newcommand{\bma}{\begin{matrix}}
\newcommand{\ema}{\end{matrix}}
\newcommand{\bpm}{\begin{pmatrix}}
\newcommand{\epm}{\end{pmatrix}}

\title{Supersymmetric Rings in Field Theory}
\author{Jose J. Blanco-Pillado {\it and} Michele Redi\\
Department of Physics and CCPP, New York University, \\
4 Washington Place, New York, NY 10003, USA\\
E-mail: \email{blanco-pillado@physics.nyu.edu}\\
E-mail: \email{redi@physics.nyu.edu}}

\preprint{}

\abstract{We study the dynamics of BPS string-like objects obtained
by lifting monopole and dyon solutions of $N=2$ Super-Yang-Mills
theory to five dimensions. We present exact traveling wave solutions
which preserve half of the supersymmetries. Upon compactification
this leads to macroscopic BPS rings in four dimensions in field
theory. Due to the fact that the strings effectively move in six
dimensions the same procedure can also be used to obtain rings in
five dimensions by using the hidden dimension.}

\keywords{Supersymmetric Theories, Solitons}

\begin{document}

\section{Introduction}

In recent years many phenomena of $D-$brane physics have found a
very similar implementation in the more conventional arena of
supersymmetric field theories so that it is natural to consider
similar constructions in this context. In this note we study
traveling wave solutions on solitonic strings and use them to build
BPS rings in field theory. Our starting point will be pure $N=2$
Super-Yang-Mills (SYM) in four dimensions. This celebrated theory
has BPS monopoles and instantons (for reviews see
\cite{harvey,tong}). By lifting these solutions to higher dimensions
one obtains BPS strings in five and six dimensions. We consider
traveling waves propagating on these strings and present the exact
field theory description of these configurations. With the explicit
wave solutions at our disposal we can then build stationary rings in
one lower dimension by compactifying the theory on a circle. The
rings appear as the lower dimensional projection of the string
profile traveling in the extra-dimension. The excitations of the 
string probe the extra dimension so in this regard these configurations 
are not solutions of the lower dimensional theory.

At first sight it might look surprising that we find supersymmetric
rings in $4D$ given the fact that a closed loop can only carry a
scalar central charge and the only BPS particle solutions of $N=2$
SYM are monopoles and dyons. The resolution to this is simply that
the rings are intrinsically higher dimensional objects. From the
four dimensional point of view they carry a monopole charge and the
extra-dimensional momentum which contribute to the same central
charge in the supersymmetric algebra. For a given monopole charge
the loop can have any size depending on the momentum flowing in the
extra-dimension. Similar arguments hold for the instanton strings in
$6D$. In $5D$ we also study non-relativistic strings obtained from
dyon solutions and their generalizations. The waves travel at a
speed smaller than the speed of light in this case.

The paper is organized as follows. In section \ref{construction} we
provide the general strategy to construct static ring solutions in
$d$ dimensions starting from an infinite string in $d+1$ dimensions.
In section \ref{AH} we present a first example of this procedure in
the Abelian Higgs model. In \ref{sigma} we comment on the
generalization of our construction to account for multiple winding
strings in the context of $\sigma-$model strings. In section
\ref{monopolestring} we consider in detail the string obtained by
lifting the monopole solution of $N=2$ SYM and present the general
traveling wave solution. Section \ref{susy} discusses the BPS
properties of the traveling waves. In \ref{dyonssec} we analyze the
dyonic strings and their excitations. In \ref{hidden} we show how
ring-like solutions can be obtained even without the presence of a
real extra-dimension when the string has extra-zero modes besides
the position. We conclude in section \ref{conclusion}. In the
appendix we generalize these results to the instanton strings in six
dimensions.

\section{Construction}
\label{construction}

In this section we describe the general method to construct the
static loop configurations (rings) from solitonic strings
propagating in one more dimension. The first step in our
construction is to obtain the field theory solutions for the
string-like objects. This can be done, for objects of different
co-dimension, by first identifying the appropriate field theories
that possess point-like solitonic solutions, such as vortices in
$2+1$ dimensions, monopoles in $3+1$, etc... and then uplift those
solutions to one more dimension by considering the field theory
configurations invariant along the extra-dimension.

In general, the solutions in the lower dimensional theory depend on
a set of integration constants. These integration constants
parameterize the massless zero modes of the $1+1$ low energy
effective action. There could be several such zero modes associated
with different aspects of the solution as its position, size,
orientation in field space. An excitation of the zero modes
corresponds to a traveling wave. For our purposes we mainly focus on
the zero modes corresponding to the position of the string. If the
object is Lorentz invariant along its extended direction, i.e its
energy momentum tensor is that of a relativistic string, geometrical
arguments imply that the zero modes corresponding to the position of
the string are governed by the standard Nambu-Goto action,
\begin{equation}
S=-T \int{d\tau d\sigma  \sqrt{-\gamma}}
\end{equation}
where $T$ denotes the tension of the string, ($\tau, \sigma$) are
the world-sheet coordinates and $\gamma$ is the determinant of the
induced metric on the string $\gamma_{ab}$, where $a$ and $b$ denote
the internal indexes on the worldsheet.

\begin{figure}
\centering\leavevmode \epsfysize=6cm \epsfbox{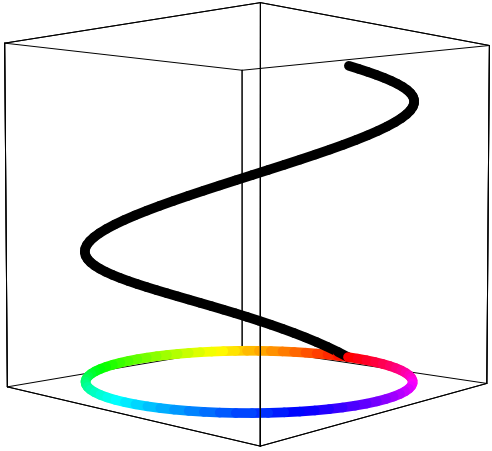}
\caption[Fig 1] {Snapshot of the string configuration and its ring
projection in $2+1$ dimensions.}
\end{figure}

Solutions of the equations of motion derived from this Nambu-Goto
action have been known for a long time, and it is straightforward to
find solutions describing traveling waves moving at the speed of
light along an infinite straight string \cite{Vilenkin}. In the
static gauge $X^0=\tau$, $X^d=\sigma$, these solutions take the
form,
\begin{eqnarray}
\label{solution-thin-wall}
X^0 &=& \tau \nonumber \\
X^i &=& \psi^i(\sigma \pm \tau) \nonumber \\
X^d &=& \sigma~.
\end{eqnarray}
The functions $\psi^i(\sigma\pm \tau)$ are arbitrary so that the
wave can have any shape provided that the excitation travels in only
one direction. These type of solutions have special interest in
string theory \cite{Callan, Dabholkar,Lunin,Cho,Mateos,BP}.

Let's now consider the significance of these solutions on a
compactified space-time of the form $M_d \times S^1$. A straight
string wrapping around the extra-dimension would be seen from
$d-$dimensions as a point particle. On the other hand, if we allow
the string to carry a wave of the position zero mode, the profile of
the string will appear to the low energy observer as a closed loop
of the string. This can be achieved in the scheme described above by
imposing that the functions $\psi^i$ have the same periodicity as
the extra dimension \footnote{More in general one can consider
strings winding $k$ times around the $S^1$ in which case the
constraint becomes $\psi^i(z)=\psi^i(z+2 k \pi R)$.},
\begin{equation}
\psi^i(z)=\psi^i(z+2 \pi R)
\end{equation}
in which case the solution (\ref{solution-thin-wall}) is also
solution of the theory on $M_d \times S^1$ where $z$ parametrizes
the angular coordinate along the circle, i.e. $0<z<2 \pi R$. The 
energy of the configuration takes the simple form,
\begin{equation}
E=T\int_0^{2 \pi R} d\sigma (1+ \vec{\psi}'^2).
\label{ngenergy}
\end{equation}
This is just the sum of the energy of the unperturbed string and the
momentum carried by the wave in the extra-dimension. Since for a
macroscopic ring the amplitude of the fluctuation must be larger
than the radius of the extra dimensions, it follows from this
equation that the energy of the configuration is dominated by the
momentum.

The fact that the lower dimensional observer detects a static loop might be
puzzling at first because loops of relativistic strings should
contract under the effect of their tension. The resolution to this
puzzle is very simple. The low energy observer would not be able to
probe the position of the string in the extra-dimensional space, but
would definitely see their effect on the motion of the string
because the effective action of the string is now modified by the
presence of the scalar field $\chi(\sigma, \tau)$ that parameterizes
the position of the string along the extra-dimensional circle,
$S^1$, namely,
\begin{equation}
\label{NG-extra-d} S_{eff} =- T \int {d^2 \xi~\sqrt{-\gamma_{d+1}}} = - T  \int
{d^2 \xi~\sqrt{-\gamma}~\sqrt{1 - \gamma^{ab} \partial_a \chi \partial_b \chi}}.
\end{equation}
This type of effective action for a string propagating in four
dimensions was originally studied in the context of dimensional
reduction by several authors \cite{Nielsen}, and it may be regarded
as a particular example of a much broader type of models where the
strings can have new degrees of freedom propagating on the
worldsheet, the superconducting string models
\cite{Witten}\footnote{Here we used the word superconducting to
refer to all the string models with extra degree of freedom even if
their currents are not coupled to any electromagnetic field.}.

These new degrees of freedom affect the mechanical properties of the
string by reducing its local effective tension thus allowing the
existence of static closed loop configurations
\cite{Davidson,Davis}\footnote{These stable loops are known in the
cosmic string literature as 'vortons'. See
\cite{Cosmic-Strings-Books} and references therein.}. The stability
of the string can be understood either from the lower dimensional
point of view by the angular momentum created by the induced current
along the string \cite{Davis} or from the higher dimensional
perspective as the angular momentum carried by the travelling wave
in the plane orthogonal to the propagation of the wave
\cite{Lunin,Mateos}.

In the rest of the paper we would like to show how solutions of the
type described above can be realized as smooth solitonic objects in
ordinary field theory. In fact we will find that these solutions
obtained in the thin-limit can be lifted to solutions of the full
field theory equations for arbitrary large fluctuations.

\subsection{Abelian-Higgs model strings.}
\label{AH}

In order to see how the construction outlined before works
in field theory, we will first consider the Abelian-Higgs model
in $3+1$ dimensions,
\begin{equation}
S_{AH}= \int{d^4x \left[|D_{\mu} \phi|^2 - {1\over 4} F_{\mu \nu}
F^{\mu \nu} -{e^2\over 2} (|\phi|^2-\eta^2)^2\right] }
\end{equation}
where, as usual, we define, $D_{\mu} \phi = (\partial_{\mu} - i e
A_{\mu})\phi$ and $ F_{\mu \nu} = \partial_{\mu}
A_{\nu}-\partial_{\nu} A_{\mu}$. The constants have been chosen so
that this action is the bosonic part of a supersymmetric theory. As
is well known this theory admits solutions which satisfy first order
Bogomol'nyi equations describing a vortex in the $x-y$ plane. These
solutions preserve half of the supersymmetries and are characterized
by the quantized magnetic flux. As shown in a nice paper by
Vachaspati and Vachaspati \cite{VV}, it is possible to find the
general solution of an arbitrary traveling wave on a straight
string. The key point of that paper was to identify the precise
modifications that one would have to introduce to the ansatz in
order to solve the full nonlinear equations of motion. They then
showed that the ansatz,
\begin{eqnarray}
\label{travelling-wave-ansatz}
\phi&=&\Phi(X,Y)\nonumber \\
A_x &=& {\cal A}_x(X,Y) \nonumber\\
A_y &=& {\cal A}_y(X,Y)\nonumber  \\
A_t &=& - \psi'_x {\cal A}_x(X,Y) -  \psi'_y {\cal A}_y(X,Y) \nonumber \\
A_z &=& \pm A_t
\end{eqnarray}
solves the equations of motion for any $\Phi$ and ${\cal A}_i$ which
satisfy the static Bogomol'nyi equations. The solution depends on
$X=x-\psi_x(z \pm t)$ and $Y=y-\psi_y(z \pm t)$ which denote the
shifted time dependent positions of the string in the $x-y$ plane
for two arbitrary functions $\psi_x$ and $\psi_y$. It can be shown by
considering the variations of the fermions that complete the
supersymmetric theory that the traveling waves preserve or break the
supersymmetries of the background depending on the direction of the
wave. We will explain this fact in section \ref{susy}.

The wave (\ref{travelling-wave-ansatz}) is the field theory
realization of the Nambu-Goto solution (\ref{solution-thin-wall}).
We can also compute the energy of this configuration at the field
theory level which agrees with (\ref{ngenergy}) showing that the
dynamics of the traveling waves is exactly captured by the
Nambu-Goto approximation.

\subsection{Multiple winding solutions. ($\sigma-$model Strings)}
\label{sigma}

In the string theory examples of Refs.
\cite{Dabholkar,Callan,Lunin}, configurations with multi-strand
strings or multiple winding strings were also considered. In
principle the same could be done for the field theory strings.
Generically if the field theory is supersymmetric there exist BPS
multi-center solutions and so there will be configurations with
independent waves for each of these strings. Unfortunately for the
gauge strings of the Abelian-Higgs model (or of SYM considered in
the following sections) the wave ansatz
(\ref{travelling-wave-ansatz}) does not generalize in an obvious way
to multi-center solutions. Here we consider the same problem in the
context of $\sigma-$model strings in $3+1$ dimensions and show that
in that case multi-center waves can easily be found.

It is well known that non-linear $\sigma-$models with K\"ahler
target space admit static string configurations
\cite{Gibbons-Comtet}. These solutions are classified by a
topological number and the energy of the configuration, in that
topological sector, is bounded by its charge. This is due to the
fact that this model can be embedded in a supersymmetric model whose
target manifold is K\"ahler. Here for simplicity we will investigate
a particularly simple model, the $CP^1$ model. The advantage of
using this simple model is that we will obtain an explicit
expression for the multiple winding string by making use of the well
known multiple center solutions but let us remark that a similar
construction will work in general. For the $CP^1$ model the action
is given by,
\begin{equation}
S = \int{d^4x~{{\partial_{\mu} \phi \partial^{\mu} {\bar \phi }}
\over {\left(1+ \phi {\bar \phi}\right)^2}}}
\end{equation}
from which we obtain the equation of motion,
\begin{equation}
\label{general-eom}
\partial_{\mu} \partial^{\mu} \phi - {{2 \bar{\phi}\, \partial_{\mu} \phi
\partial^{\mu} \phi}\over  {1+ \phi {\bar \phi}}} =0.
\end{equation}
We are interested in static string-like solutions of this equation
where the fields only depend on the $x$ and $y$ coordinates. The
equation above is satisfied by any holomorphic function $\phi(x+i
y)$. An isolated string of unit topological charge centered at zero
corresponds to,
\begin{equation}
\phi(x+i y) = {1 \over {x+i y}}~.
\end{equation}
Much in the same way as in the Abelian-Higgs model we obtain a
traveling wave solution by taking,
\begin{equation}
\phi(x,y) = {1 \over {((x-\psi_x(z \pm t)+ i (y-\psi_y(z \pm t))}}~.
\end{equation}
This configuration is therefore another field theory realization of
the string wave characterized, in the thin wall approximation, by
the solution (\ref{solution-thin-wall}). The simplicity of the
$CP^1$ model allows us to write simple solutions of higher
topological charge, such as,

\begin{figure}
\centering\leavevmode \epsfysize=8cm \epsfbox{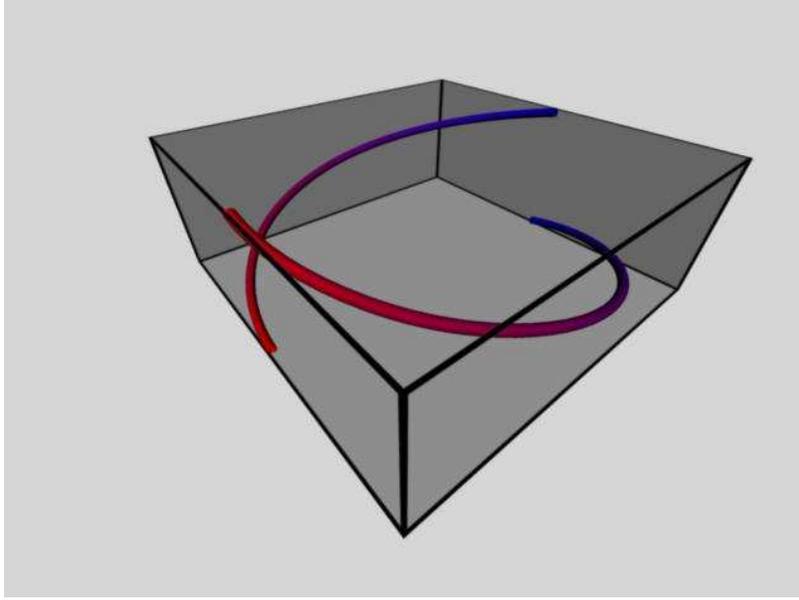} \caption[Fig
1] {Snapshot of the multiwinding string configuration.}
\end{figure}

\begin{equation}
\label{multicenter} \phi_2(x,y) = {1 \over {((x-1) + i y))((x+1) + i
y)}}
\end{equation}
which represents two strings of unit charge located at $(x=1,0)$ and
$(x=-1,0)$. We can now wonder what kind of configurations we will
obtain upon compactification by exciting traveling waves on the
different strings. It is clear that we could get a collection of
``disconnected'' rings in $2+1$ dimensions, each of which associated
with a different string. There is however a more interesting
possibility of constructing a single ring by imposing the correct
periodicity on each of the strings in such a way that the end result
becomes a single string of higher winding along the extra dimension.
A simple example of this can be obtained by using the previous
multicenter solution (\ref{multicenter}) and taking for simplicity a
sinusoidal traveling wave,
\begin{equation}
\phi_2(x,y) = {1 \over {((x- \cos(z \pm t)) + i
(y-\sin(z \pm t)))((x+ \cos(z \pm t)) + i (y+\sin(z \pm t)))}}~.
\end{equation}
This configuration describes a couple of helicoidal strings
propagating along the $z$ direction. By requiring the $z$ direction
to be compact with period $\pi$ we can identify the previous
solution as a single string winding two times along the extra
dimension before closing on itself (See Fig. (2)). On the other
hand, it is also possible to find this kind of string state as a
solution of the effective action (\ref{NG-extra-d}),
\begin{eqnarray}
\label{solution-thin}
X^0 &=& \tau \nonumber \\
X^1 &=& \cos(\sigma \pm \tau) \\
X^2 &=& \sin(\sigma \pm \tau) \\
\chi &=& \nonumber \sigma
\end{eqnarray}
where $0 < \sigma < 2 \pi$ and therefore, the field $\chi$  winds
two times as we move along string loop. It is straightforward to
extend the solution found above to more general wave forms as well
as to higher winding number.

We would also like to note that, as in many of the cases we will
talk about in this paper, there are other zero modes on the low
energy effective theory of these strings. Exciting some of those
along the compactified direction will lead to different type of
ring-like configurations in the lower dimensional space-time. An
example of such situation is the Q-lump solution in the $CP^1$ model
\cite{Leese, Abraham}.

\section{Monopole Strings in $5D$}
\label{monopolestring}

Let us now see how we can obtain ring solutions in $3+1$ dimensions.
Following the procedure described in the previous section, we should
first find a model that possesses as part of its spectrum
string-like objects in $4+1$ dimensions. The simplest supersymmetric
theory of this kind corresponds to $SU(N)$ Super-Yang-Mills theory
in $4+1$ dimensions. The bosonic part of the action is given by,
\begin{equation}
\label{YMH} S_{YMH}= \int{ d^5 x~ Tr \left[-{1 \over 2} F_{M N} F^{M
N} +  D_{M} \phi D^M \phi \right]}
\end{equation}
with
\begin{eqnarray}
F_{MN}&=& \partial_M A_N - \partial_N A_M + i e [A_{M},
A_{N}]\nonumber\\
D_M \phi &=& \partial_M \phi +i e [A_M,\phi],
\end{eqnarray}
where $\phi$ and $A_M$ are valued in the algebra of $SU(N)$. Upon KK
reduction to four dimensions the theory reduces to pure $N=2$ SYM.
As well known, this theory admits supersymmetric monopole solutions.
These configurations correspond to solutions of the first order
equations,
\begin{equation}
B_i=\pm D_i \phi \label{bogo}
\end{equation}
where $B_i=-1/2 \epsilon_{ijk} F_{jk}$. The solutions saturate a
Bogomol'nyi bound on the energy where the energy is equal to the
magnetic charge in appropriate units. For simplicity of notation in
the rest of the paper we will restrict our analysis to the case
where the gauge group is $SU(2)$ but the generalization to $SU(N)$
is immediate. For the case of monopole charge one, the explicit
solution of the Bogomol'nyi equations is given by,
\begin{eqnarray}
\phi &=& \Phi = {{x^a \sigma^a} \over {2 e r^2}} H(v e r) \nonumber \\
A_i &=& {\cal A}_i = - \epsilon_{aij}  {{x^j \sigma^a} \over {2 e r^2}} (K(v e r) -1)\nonumber  \\
A_0 &=& A_4 = 0~, \label{onemonopole}
\end{eqnarray}
where $r$ is the $4D$ radial direction, $v$ the asymptotic value
of $\phi$, $\sigma^a$ are Pauli matrices and
\begin{eqnarray}
H(y) &=& y \coth(y) - 1\nonumber  \\
K(y) &=& {y \over {\sinh(y)}}~. \label{functions}
\end{eqnarray}
The monopole solutions written above obviously solve the full five
dimensional equations of motion since nothing depends on the fifth
coordinate. In five dimensions these solutions become string-like
configurations where each section of the string is given by the
monopole solution. Their tension is given by,
\begin{equation}
T=\frac {4 \pi v} e,
\end{equation}
where to agree with the $5D$ normalizations $v$ and $e$ have mass
dimensions $3/2$ and $-1/2$ respectively.

Having found smooth strings in $5D$ we now wish to consider waves
traveling on these objects, i.e. excitations of the zero modes
living on the string. In particular to construct rings we are
interested in the waves of the position of the string. The five
dimensional equations of motion that follow from \ref{YMH} are given
by,
\begin{eqnarray}
\label{monopole-eom}
D_M F^{MN} &=&i e \left[\phi, D^N \phi\right]~,\nonumber \\
D^M D_M \phi &=& 0.
\end{eqnarray}
In order to find the exact field theory description for the
traveling waves on the monopole-string we closely follow the example
of the Abelian-Higgs model presented previously. Promoting the
string position to a light-like field the equation of motion for the
scalar and the components of the gauge field equations transverse to
the string are automatically satisfied as long as $A_0=\pm A_4$.
These can be chosen so that also the 0 and 4 equations for the gauge
field are solved. In fact, despite of the non-abelian nature of the
monopoles, the straightforward generalization of the abelian case
works for the monopole-strings as well. The most general solution
representing a transverse traveling wave moving at the speed of
light along the $x^4$ direction is given by,
\begin{eqnarray}
\phi &=& \Phi(X,Y,Z) \nonumber \\
A_i &=& {\cal A}_i (X,Y,Z)~~~~~~~~~~~~~~i=1,2,3, \nonumber \\
A_0 &=& -\sum_{i=1}^3 \psi_i' {\cal A}_i \nonumber \\
A_4 &=& \pm~A_0 \label{monopolewave}
\end{eqnarray}
where $\bold \psi = \psi_i(x^4 \pm x^0)$ is the three dimensional
vector that parameterizes the arbitrary transverse perturbation, and
$(X,Y,Z)= (x_i-\psi_i(x^4\pm x^0))$ are the shifted position of the
string core due to the presence of the fluctuation. This solution
actually holds not only for the one monopole solution
(\ref{onemonopole}) but for any solution of the Bogomol'nyi
equations (\ref{bogo}). In the more general case of the multi-center
solutions there will be traveling waves associated with the motion
of each individual string. The solution (\ref{monopolewave})
describes the excitation corresponding to the shift of the center of
mass of the system.

Beside the zero modes corresponding to the position, the
monopole-string has yet another bosonic collective coordinate $\chi$
associated to the global $U(1)$ transformation (see \cite{harvey}
for the details). There are traveling waves solutions for this mode
as well, as for any other zero mode that lives on the string such as
fermionic zero modes required by supersymmetry\footnote{Stable
monopole string rings may also be obtained by exciting fermionic
waves on the string in a similar way to the configurations recently
proposed in \cite{BPI-2}.}. The traveling wave solution is given in
this case by,
\begin{eqnarray}
\phi &=& \Phi\nonumber\\
A_0&=&\chi' ~\Phi \nonumber \\
A_i&=& {\cal A}_i  \nonumber \\
A_4&=&\pm A_0 \label{dyonwave}
\end{eqnarray}
where $\chi$ is an arbitrary function of $(x^4\pm x^0)$. The
$\chi$-wave can be added to the waves of the positions giving the
most general bosonic traveling wave on the monopole string which
depends on four arbitrary functions. From this solution we can see
that $\chi$ appears on the same footing as the other fluctuations,
if we interpret the scalar field as the fifth component of a gauge
field in six dimensions. Actually this is not an accident and as we
will see in section \ref{dyonssec} it follows from the fact the
monopole string effectively moves in six dimensions. As a special
case of (\ref{dyonwave}), we can take $\chi$ to be linear in
$(x^4\pm x^0)$. With this choice we obtain a static configuration
translationally invariant along the world-volume (this would not be
possible for the wave of the position). Since these solutions do not
depend on on $x_4$ they are also a solutions of SYM in $4D$; in fact
they are contained within the general solutions found in
\cite{dadda}. From the $5D$ point of view these solutions carry
linear momentum and reproduce the ones found in \cite{Bak}. Note,
however that our solutions in \ref{dyonwave} are more general than
the ones obtained in the literature since they hold for any function
$\chi(x^4 \pm x^0)$, but those depend explicitly on $x^4$ so they
are not solutions of the lower dimensional theory.

\subsection{Supersymmetry}
\label{susy}

We now examine the traveling wave solutions found in the previous
section from the supersymmetry point of view (see also
\cite{kimlee}). As we will show most of the properties can be
derived in a model independent way from the supersymmetry algebra.
Very similar arguments hold for the Abelian-Higgs model vortices and
$\sigma-$model strings considered in the section \ref{construction}.

The supersymmetric extension of the action (\ref{YMH}) is invariant
under $N=1$ supersymmetry in $5D$ which corresponds to $N=2$
supersymmetry in $4D$. The variation of the fermions under
supersymmetry is,
\begin{equation}
\delta \psi=\left(\gamma^{MN}F_{MN}-\gamma^M D_M \phi\right)\epsilon
\label{variations}
\end{equation}
where $\epsilon$ is a four component Dirac spinor. The straight
string is just the lift to $5D$ of the $4D$ BPS monopoles where all
the fields are independent from the fifth coordinate. It follows
immediately that this state is supersymmetric. What is perhaps less
obvious but still true is that the waves traveling at the speed of
light on the string can also be BPS. Using the explicit expressions
(\ref{monopolewave}) in the fermion variations (\ref{variations}),
it is possible to check that these solutions are one half BPS as the
background if the wave travels in the direction specified by the
monopole charge (i.e. upward for positive monopole charge in our
conventions). Waves traveling in the opposite direction break all
the supersymmetries instead. The ring configuration are obtained as
a special case of traveling wave solutions so they are also
supersymmetric as long as the wave travels in the appropriate
direction. From the lower dimensional point of view these objects
carry a scalar central charge as the magnetic monopoles. More
precisely the central charge contains two contributions, one is the
magnetic charge and the other the momentum flowing in the
extra-dimension.

We would now like to show that the BPS properties of the traveling
wave are a direct consequence of the supersymmetry algebra in $5D$.
This algebra is formally identical to the $N=2$ supersymmetry
algebra in $4D$ which in Weyl notation is given by,
\begin{eqnarray}
\left\{Q_{\alpha}^A,\bar{Q}_{\dot{\alpha}B}\right\}&=&\sigma^{\mu}_{\alpha\dot{\alpha}}
P_{\mu}\delta^A_B \nonumber \\
\left\{Q_{\alpha}^A,Q_{\beta}^B\right\}&=&\epsilon_{\alpha\beta}
\epsilon^{AB} Z\label{algebra2}.
\end{eqnarray}
In $4D$ $Z$ is a complex central charge carried by BPS point-like
objects such as monopoles and dyons. With this parametrization, in
$5D$ the imaginary part of $Z$ represents the fifth component of the
momentum so that the algebra has single real central charge. A
string can both carry a scalar central charge or a vectorial central
charge (the algebra also contains a triplet of 2-index central
charges but this cannot be carried by a string). Considering the
Lorentz transformations of the charges it follows that an object
Lorentz invariant along the world-volume direction can only carry a
vectorial central charge. Since monopole strings are manifestly
Lorentz invariant it follows that the monopole charge of the string
enters in the algebra as ``momentum'' since $P^M$ is the only
vectorial charge in the $5D$ algebra. Explicitly the topological
charge of the string in $5D$ is the vector,
\begin{equation}
P^M_{mag}=\epsilon^{0MNPQ} \int d^4x\,Tr[F_{NP}\, D_Q \phi].
\label{magneticcharge}
\end{equation}
For an object carrying $P^4$ charge, the algebra (\ref{algebra2})
implies the bound on the energy,
\begin{equation}
E \ge |P^4|. \label{bpsbound}
\end{equation}
The formula is identical to the one of massless particles but now to
$P^M$ contributes the linear momentum as well as the magnetic charge
(\ref{magneticcharge}). When the bound is saturated, as for the
monopole strings, the object preserves half of the supersymmetries.
Let us now add fluctuations of the zero modes living on the string
moving in one direction. Since the excitation moves at the speed of
light in space-time its energy must be equal to momentum. To see
this more explicitly we can compute the energy momentum tensor,
\begin{equation}
T_{MN}=Tr \left[ \frac 1 4 g_{MN} F_{PQ}F^{PQ}-  F_{MP}F_N^P-\frac
1 2 g_{MN}(D_P \phi)^2+ D_M \phi D_N \phi \right] \label{emt}.
\end{equation}
Using the solution (\ref{monopolewave}) one finds,
\begin{equation}
\delta T_{00}=|T_{04}|=Tr \left(\sum_{i=1}^3 \psi'_i D_i
\phi \right)+Tr\left( \sum_{i,j,k=1}^3 F_{ij} F_{ik}\psi'_k \psi'_j\right)
\end{equation}
where $\delta T_{00}$ denotes the difference of energy densities
between the excited and the background string solutions (in the
previous expression all the fields are evaluated at the shifted
point $(x_i-\psi(x^4\pm x^0))$. By integrating over the transverse
directions the total energy reduces exactly to eq. (\ref{ngenergy})
obtained from the Nambu-Goto action. Therefore the total charge
appearing in the algebra is given by,
\begin{equation}
P^4=P^4_{mag}+P^4_{lin}\nonumber
\end{equation}
where,
\begin{eqnarray}
P^4_{mag}&=&2\pi R \frac {4 \pi v} e \nonumber \\
P^4_{lin}&=&\pm \frac {4 \pi v} e \int_0^{2 \pi R} dx^4 \sum_i
\psi_i'^2
\end{eqnarray}
The total energy is given by $E=|P^4_{mag}|+|P^4_{lin}|$. From the
BPS bound (\ref{bpsbound}) we derive that the vibrating string is
BPS only if $P^4_{mag}$ and $P^4_{lin}$ have the same sign. When
this condition is satisfied the traveling wave leaves unbroken the
same supersymmetries of the background, it is invariant under four
supercharges. Note that these arguments do not depend on the type of
fluctuation as long as it travels at the speed light in the
direction specified by the orientation of the string. It is useful
to consider these facts from point of view of the 1+1 dimensional
world-volume theory. Consider the background string solution.
Projecting the $5D$ algebra on the subspace of the unbroken
generators one finds that the unbroken supersymmetries form a
$(0,4)$ two dimensional superalgebra. This means that the
supersymmetries act only on the movers in one direction, the string
is chiral. As a consequence a massless excitation in the opposite
direction leaves unbroken four supersymmetries. Let us consider the
fermions. The zero modes of the monopole solution are four bosonic
coordinates and four fermions which have a definite chirality. Upon
lifting the solution to five dimensions there will be four chiral
fermions moving in only \emph{one} direction and four bosons moving
in \emph{both} directions. The fermions are the goldstinos for
partial breaking of supersymmetry and so they transform in the
bosons under the action of the unbroken supersymmetries. As a
consequence a wave of the fermions breaks all the supersymmetries.
If however there are additional fermions on the string moving in the
appropriate direction a wave of the fermions will also be BPS.

The scenario with fermions moving in both directions is
automatically realized in $N=4$ SYM. The previous analysis can be
extended to this case with minor modifications (the algebra now
contains a triplet vectorial central charge which is carried by the
string). The background monopole string solution is identical to the
one considered before. In this case the unbroken supersymmetries
form a $(4,4)$ two dimensional superalgebra. On the monopole string
there are four fermions traveling in each direction which match the
number of bosonic zero modes so the string is non-chiral. From this
we can conclude that any wave is a quarter BPS state preserving the
supercharges with opposite chirality with respect to the wave.
\section{Non Relativistic Strings} 
\label{dyonssec}
The $5D$ SYM admits several other BPS strings. The properties of these
objects however are rather different from the monopole strings since
as we will show generically they break Lorentz invariance along the
string.

One type of string-like object can be obtained by lifting the dyon
solution of $4D$ SYM to $5D$. As shown by Mueller long ago
\cite{Muller}, the dyon solution is related to the monopole solution
by a boost. For the monopole string which is supported by $\phi$
this can be achieved interpreting $\phi$ as the fifth component  of
a gauge field in six dimensions and boosting in that
direction\footnote{This procedure can be made explicit by lifting
the monopole string to $6D$ SYM where it becomes a supersymmetric
(but non-relativistic) membrane.}. The solution is given by,
\begin{eqnarray}
\phi&=& \Phi \left(\frac {x^a}{\cosh \theta}\right) \nonumber \\
A_0&=&\tanh \theta\,  \Phi\left(\frac {x^a}{\cosh \theta}\right) \nonumber \\
A_i&=&\frac 1 {\cosh \theta} {\cal A}_i \left(\frac {x^a} {\cosh \theta}\right)\nonumber \\
A_4&=& 0 \label{dyon}
\end{eqnarray}
where $\Phi$ and ${\cal A}_i$ are the monopole string solution
(\ref{onemonopole}). Note that the fields have been rescaled so that
they approach the same vacuum. The energy per unit length is given
by the dyon formula,
\begin{equation}
\rho=v \sqrt{Q_e^2+Q_m^2} \label{dyonmass}
\end{equation}
where,
\begin{equation}
Q_e=Q_m \sinh \theta.
\end{equation}
The energy density is identical to the one for the $(p,q)$ strings
of type IIB string theory. In fact the formula just depends on the
structure of the supersymmetry algebra which turns out to be similar
in the two cases. There is however an important difference. From the
point of view of the algebra in $5D$ the magnetic charge is the
``momentum'' charge while the electric charge contributes to the
scalar central charge of the supersymmetric algebra,
\begin{equation}
X_{el}=v\, Q_e=2 \int d^4x~Tr[F_{0i}\,D_i \phi].
\end{equation}
The presence of the scalar central charge breaks Lorentz invariance
along the string, i.e. the string is not relativistic. This also
follows immediately from the solution since $A_0$ is different from
zero.

Broken Lorentz invariance drastically alters the properties of the
zero modes living on the string. In particular the zero modes will
not travel at the speed of light. To see this explicitly one can
compute the energy momentum tensor (\ref{emt}) for the dyon string.
From the ratio of the pressure $T_{44}$ to the energy density
$T_{00}$ (all other components of $T_{MN}$ are equal to zero), we
learn that the velocity of propagation on the string is given by,
\begin{equation}
v_s=\frac {Q_m}{\sqrt{Q_e^2+Q_m^2}}=\frac 1 {\cosh \theta}
\label{dyonvelocity}
\end{equation}
This formula can be directly obtained from the supersymmetry algebra
by noticing that the only contribution to the tension arises from
vectorial central charges.

The field theory waves on the dyonic strings cannot be easily found
as for the monopole strings. The trick of boosting in the $\phi$
direction does not work in this case because for a traveling wave it
would introduce a dependence on the extra coordinate. In fact since
the speed of the traveling waves is smaller than the speed of light
it is not even obvious at first that these excitations can be
supersymmetric. We can however argue in the following way. As argued
by Tong \cite{tong} the effective action for the monopole strings is
the Nambu-Goto action of a string moving in one extra dimension
$\chi$ which is compactified on a circle of radius $1/(ev)$,
\begin{equation}
S=- T~\int d^2 \sigma \sqrt{-\gamma_{6}}
\end{equation}
where $\gamma_{6}$ is the determinant of the pull back of the six
dimensional flat metric where $X_5=\chi$. This action can be derived
by considering the exact spectrum of BPS dyons in $4D$. A very
important fact is that the action is Lorentz invariant in six
dimensions, it effectively describes a relativistic string in $6D$.
At least for the maximally supersymmetric case this string can be
identified with the strings of the $(2,0)$ superconformal field
theory in $6D$ which is related to $5D$ SYM by compactification (see
\cite{tong,seiberg} and Refs. therein). From the Nambu-Goto point of
view the dyon string corresponds to the monopole string moving with
constant speed in the $\chi$ direction, $\chi=\beta t$. The constant
$\beta$ is related to the electric charge by,
\begin{equation}
\beta=\frac {Q_e}{\sqrt{Q_e^2+Q_m^2}}
\end{equation}
We can now consider waves propagating on the dyon background. From
the Nambu-Goto action one finds that there are waves of arbitrary
shape traveling at the reduced velocity (\ref{dyonvelocity}),
\begin{eqnarray}
X^0 &=& \tau \nonumber \\
X^i &=& \psi^i(\sigma \pm v_s \tau) \nonumber \\
X^4 &=& \sigma\nonumber\\
X^5 &=&\beta\, \tau. \label{ngtong}
\end{eqnarray}
This solution is a boost along the $\chi$ direction of the wave on
the monopole string. The reduced velocity can then be interpreted as
the red-shift due to the motion of the center of mass of the string
in the hidden dimension. Since the effective action describes a
supersymmetric string in six dimensions it is also clear that the
traveling waves are supersymmetric (as long as they travel in the
appropriate direction), because they are just the boost of the
supersymmetric waves on the monopole strings. The existence of
supersymmetric traveling wave solutions at the level of the
Nambu-Goto action strongly suggest that these solutions will be
supersymmetric in the full field theory realization.

\subsection{Hidden Dimension}
\label{hidden}

Once we realize that the monopole string is a six dimensional
relativistic string many other solutions are immediately available
at the level of the Nambu-Goto action. Many of these solutions have
a field theory correspondent. Starting from the monopole string in
six dimensions,
\begin{eqnarray}
X^0 &=& \tau \nonumber \\
X^i &=& 0 \nonumber \\
X^4 &=& \sigma\nonumber\\
X^5&=& 0, \label{monopoleng}
\end{eqnarray}
we can obtain new strings in $5D$ by boosting in the $X^5$ direction
and by rotating in the plane $X^4-X^5$. The most general solution of
this kind is given by,
\begin{eqnarray}
X^0 &=& \tau \nonumber \\
X^i &=& 0 \nonumber \\
X^4 &=& \sigma\nonumber\\
X^5&=& \frac{\tanh \theta} {\cos \alpha}\, \tau-\tan \alpha\,
\sigma, \label{namburb}
\end{eqnarray}
where we have reparametrized the world-sheet coordinates by choosing
the static gauge along $X^4$. The boost and rotation has a direct
analog in field theory. Treating again $\phi$ as the fifth component
of a gauge field in $6D$ we can obtain static strings in $5D$ by
performing the same boost and rotation as in the Nambu-Goto case.
The solution obtained by a rotation for example corresponds to
lifting a $4D$ monopole solution which is supported by a linear
combination of $\phi$ and $A_5$ while the boost corresponds to the
dyon considered before. In order to find the field theory
description of the solutions (\ref{namburb}) we need however to
require that they asymptote to the same vacuum (determined by the
value of the scalar $\phi$ at infinity) as the monopole string.
Following \cite{Muller}, this can be achieved by rescaling of the
coordinates. The general solution is then given by,
\begin{eqnarray}
\phi&=& \Phi\left(\frac {x^a}{\cosh \theta \cos\alpha}\right) \nonumber \\
A_0&=&\frac {\tanh \theta}{\cos\alpha}~\Phi\left(\frac {x^a}{\cosh \theta \cos \alpha}\right) \nonumber \\
A_i&=&\frac 1 {\cosh \theta \cos\alpha}~{\cal A}_i \left(\frac {x^a} {\cosh \theta \cos\alpha}\right)\nonumber \\
A_4&=& \tan \alpha~\Phi\left(\frac {x^a}{\cosh \theta
\cos\alpha}\right). \label{dyon-2}
\end{eqnarray}
These strings have the same energy density and pressure as the one
obtained from the Nambu-Goto solution (\ref{namburb}) showing that
they are the field theory realization of those solutions. The
charges carried by these strings are in general,
\begin{eqnarray}
P^4_{mag}&=&-\epsilon^{0ijk4}\int d^4x\, Tr[F_{ij}\, D_k \phi]\nonumber \\
X_{inst}&=&-\epsilon^{0ijk4}\int d^4x\, Tr[F_{ij}\, D_k A_4]\nonumber \\
P^4_{el}&=& 2\int d^4 x\, Tr[F_{0i}\,D_i A^4] \nonumber \\
X_{el}&=& 2\int d^4x\, Tr[F_{0i}\, D_i \phi] \nonumber \\
P^4_{lin}&=&\int d^4 x\, T_0^4 \label{generalcharges}
\end{eqnarray}
where $X$ and $P^4$ refer to the fact that they appear in the
supersymmetric algebra as scalar or vector charge. Note that all the solutions
(\ref{dyon-2}) have magnetic monopole charge one. It is easy to see
that a rotation of the monopole string induces an instantonic charge
$X_{inst}$ while a boost generates $X_{el}$. The composition of
boosts and rotations generates also linear momentum so that the
general solution (\ref{dyon-2}) carry all the above charges. In the
Nambu-Goto description these charges can be identified with the
winding and momentum of the string. For example winding in the fifth
dimension corresponds to instanton charge while momentum corresponds
to electric charge.

For each of the strings (\ref{dyon-2}) there are also traveling wave
solutions. At the Nambu-Goto level these solutions are again the
boost of the traveling waves on the monopole string. One finds that
in this case the velocity of propagation is given,
\begin{equation}
v_s=\frac {\cos\alpha} {\cosh\theta}\mp  \sin\alpha \,\tanh \theta
\end{equation}
which agrees with the field theory computations. The $\mp$ refers to
the velocity of the perturbations in opposite directions which can
be different due to the fact that in general the strings are not
relativistic and carry momentum. Being Lorentz transformations of
supersymmetric solutions these waves are supersymmetric at the
Nambu-Goto level so we expect this property to continue in the
field theory solutions.

If we take seriously the Nambu-Goto action we can also find
point-like solution in five dimensions,
\begin{eqnarray}
X^0 &=& \tau \nonumber \\
X^i &=& 0 \nonumber \\
X^4 &=& 0\nonumber\\
X^5 &=& k~\sigma
\end{eqnarray}
which describes a strings winding around the hidden dimension $X^5$.
Since the radius of this dimension is $1/(e v)$ one finds that the
energy of this configuration is given by,
\cite{tong},
\begin{equation}
M=k~\left({{2\pi} \over {e v}}\right)T  = \frac {8\pi^2 k}{e^2}
\end{equation}
where $k$ is the number of windings. Remarkably this formula exactly
reproduces the mass of the instanton particles of the theory despite
the fact the solution has no string structure in $5D$. These
solutions are singular in field theory but can be considered as a
limit of the dyonic instantons of \cite{Lambert} where the electric
charge goes to zero. The $W$ bosons on the other hand can be
identified with the KK modes of the theory \cite{seiberg}. This
suggests that the Nambu-Goto description might be exact for the BPS
objects of the theory.

Following the same procedure as in the previous sections we obtain
rings in $5D$ by exciting waves of the position now traveling in the
hidden dimension. For example a circular ring lying in the $X^1-X^2$
plane is represented by,
\begin{eqnarray}
X^0 &=& \tau \nonumber \\
X^1 &=& R\, \sin\left(n\, e\, v\,(\sigma \pm  \frac {\tau} k )\right) \nonumber \\
X^2 &=& R\, \cos\left(n\, e\, v\,(\sigma \pm  \frac {\tau} k )\right) \nonumber \\
X^5 &=&k\, \sigma\, , \label{dyonicinst}
\end{eqnarray}
where $n$ is the number of times the projection of the string in
$5D$ goes around the circle in the $X^1-X^2$ plane. The mass of the
loop is given by,
\begin{equation}
M=\frac {8\pi^2 k}{e^2}+\frac{8 \pi^2 n^2 v^2 R^2} k,
\end{equation}
where the two contributions arise from the winding and the momentum
flowing in the internal dimension which correspond to the electric
and instantonic charges in (\ref{generalcharges}). These are
precisely the same charges carried by the dyonic instantons of SYM
discussed by \cite{Lambert,zamaklar,Bak,Kim}. These solutions, which
are closely related to supertubes in string theory
\cite{Supertubes,townsendfield,Hyakutake}, carry instantonic and electric
charge and their mass is given by,
\begin{equation}
M=X_{inst}\pm X_{el}
\end{equation}
which saturates the BPS bound (for a recent discussion see
\cite{kimlee}). We can therefore identify the $5D$ rings as some
kind of dyonic instantons. To show this in more detail we can look
at a small element of the loop (\ref{dyonicinst}),
\begin{eqnarray}
X^0 &=& \tau \nonumber \\
X^1 &=& \sigma \nonumber \\
X^5 &=& {{k}\over{R~n~e~v}}\sigma\,\pm \tau. \label{ringelement}
\end{eqnarray}
This has the form of the solution (\ref{namburb}) which represents
an excitation of the monopole-string we started from. We are then
lead to interpret the dyonic instantons in field theory as the
Nambu-Goto loops we have been discussing where the excitations along
the string allow the possibility of having static string loops.

We should note that, since the thickness of the monopole string in
comparable to the size of the hidden dimension, the Nambu-Goto
description could break down. On the other hand, from the field
theory realization (\ref{dyon-2}), with the parameters given by
(\ref{ringelement}), one can see that for a loop of radius $R$ the
thickness of the string constituent is given by,
\begin{equation}
\delta \sim {{n R}\over {k}},
\end{equation}
which suggests that the description in terms of the Nambu-Goto
solutions would be reliable for $k/n>>1$. In fact, already for $k=2$
it was shown by \cite{kimlee} that dyonic instantons with
string-like structure can be build. For the case $k=1$ the only
dyonic instantons known have spherical symmetry \cite{Lambert}, so
they cannot be immediately identified with the ring solutions of the
Nambu-Goto action. It is however possible that, due to the BPS
properties, even these solutions might be captured by the $6D$
Nambu-Goto action.

\section{Conclusions}
\label{conclusion}

In general, any relativistic string admits wave solutions of
arbitrary shape moving at the speed of light in space-time. These
solutions can be used to construct stationary rings in one lower
dimension since by choosing appropriately the periodicity of the
fluctuations, the traveling waves are also solutions of the theory
compactified on a circle. The ring is then the projection of the
string profile on the lower dimensional space. While the low energy
observer will only detect a static ring these solutions are truly
higher dimensional since they involve excitations of the modes of
the Kaluza-Klein tower.

In this paper we presented field theoretic realizations of the
rings. In particular we considered the solitonic strings obtained by
lifting the monopole solutions of $N=2$ SYM in $4D$ to five
dimensions (similar results hold for the instantonic strings in
$6D$, see appendix). Generalizing previous work in the context of
the Abelian Higgs model in $4D$ \cite{VV}, we have derived the exact
traveling wave solutions on these objects. These waves preserve one
half of the supersymmetries if the wave travels in the direction
determined by the charge of the monopole/instanton string. Therefore
the rings are also BPS and can be thought as some kind of blown up
monopoles.

The monopole strings studied in this paper are particularly rich
because they effectively move in six dimensions. The presence of
this extra degree of freedom allows us to construct rings in five
dimensions in an analogous way to superconducting string models and
to supertubes in string theory. Moreover a constant motion in the
hidden dimension induces electric charge on the string. The strings
so obtained are just the lift of the dyon solution to $5D$ and they
are supersymmetric but not relativistic. It would be very
interesting to find a model with similar properties in four
dimensions.

\acknowledgments We would like to thank Gia Dvali, Ken Olum, David
Tong and Tanmay Vachaspati for useful discussions and David Caeiro
for technical assistance producing the figures of this paper. J.J.B-P is
supported by a James Arthur Fellowship at NYU. The work of M.R. was
supported by the NSF grant PHY-0245068.

\appendix
\section{Instanton Strings}
\label{instanton}

In this appendix we briefly consider the generalization of this work
to the six dimensional case. In $4D$ beside supersymmetric monopoles
and dyons, $N=2$ SYM on the Coulomb branch also admits BPS instanton
solutions. Many of the arguments used for the monopole strings can
be repeated in this case as well. If we lift the theory to six
dimensions the instantons become BPS strings invariant under four
supersymmetries. In $6D$ the theory contains only a gauge field and
a Dirac spinor in the adjoint representation of the gauge group so
the lift of the $4D$ instanton is unique. The instanton string
corresponds to a solution of the self-dual equations,
\begin{eqnarray}
A_0&=&A_6=0 \nonumber \\ F_{ij}&=&\pm \star
F_{ij},~~~~~~~~~~~~~i,j=1,4 \label{selfdual}
\end{eqnarray}
i.e. its transverse slices are identical to the instanton solution
in $4D$. For the simplest case the solution to this equations can be
written in singular gauge as,
\begin{equation}
A_i=\frac {\rho^2 x_i} {r^2(r^2+\rho^2)}\eta^a_{ij} \sigma^i
\end{equation}
where $\eta^a$ is a triplet of $4 \times 4$ matrixes (see \cite{tong}).

This one instanton solution contains eight bosonic zero modes. Four
of them are the position of the instanton-string, $\rho$ determines
its size and the remaining three modes correspond to the global
gauge transformations (for the case of $SU(2)$). We now want to
consider a wave propagating on the string. The string is Lorentz
invariant so that the waves of the zero modes will propagate at the
speed of light. As usual to find traveling waves we promote the
integration constants of the static background to light-like fields.
As for the monopoles strings the exact traveling wave solution has
the form,\footnote{Traveling waves on instanton strings have also
been discussed in \cite{Eto}.}
\begin{eqnarray}
A_i &=& {\cal A}_i (X,Y,Z,T) \nonumber \\
A_0 &=& -\sum_{i=1}^4 \psi'_i {\cal A}^a_i \nonumber \\
A_5 &=& \pm~A_0~.
\end{eqnarray}
We can also build traveling wave solutions using the other zero
modes living on the string. This is particularly simple for the size
zero mode. In fact promoting $\rho$ to an arbitrary function
$\rho(x^5 \pm x^0)$ the equations of motion remain satisfied.

One can also excite the internal degrees of freedom of the $SU(2)$
symmetry and based on the experience with the monopole strings these
would also affect the mechanical properties of the strings,
rendering them non-relativistic. In this regard it is possible that
there are also stable static loops of the field theory instantonic
strings much in the same way as the ones explained in section
\ref{dyonssec} for the monopole strings where the $U(1)$ zero mode
is embedded in the $SU(2)$ case at hand. In fact, it can be shown
that there are stable string configurations at the Nambu-Goto level
when the internal manifold of the spacetime explored by the string
is a $S^2$ sphere instead of a circle \cite{BPI-1}. These solutions
however will likely not be supersymmetric.

The solutions found above are all BPS as can be checked from the
explicit supersymmetry variations. In $6D$ the minimal SUSY algebra
contains a vector charge $P_M$ and a triplet of tensorial central
charges $Z_{MN}^i$. A static straight string pointing in the fifth
direction carries the instanton charge which appears as fifth
component of the momentum in the algebra. The solution is
necessarily relativistic since there are no scalar central charges
that can be carried by the string. By looking at the algebra of the
unbroken supersymmetries one can see that as in the monopole case
these form a $(4,0)$ superalgebra in $4D$. Therefore, also in this
case only waves traveling in one direction (determined by the sign
of the instanton charge) will be supersymmetric.


\end{document}